\documentstyle[12pt]{article}

\begin{document}
\setcounter{page}{1} \pagestyle{plain} \vspace{5cm}
\begin{center}
\Large{\bf Crossing of the Phantom Divided Barrier with Lorentz Invariance Violating Fields }\\
\small
\vspace{1cm} {\bf S. Davood Sadatian}\quad\ and \quad {\bf Kourosh Nozari}\\
\vspace{0.5cm} {\it Department of Physics,
Faculty of Basic Sciences,\\
University of Mazandaran,\\
P. O. Box 47416-1467, Babolsar, IRAN\\
d.sadatian@umz.ac.ir,\, knozari@umz.ac.ir}
\end{center}
\vspace{1.5cm}
\begin{abstract}
We study possible crossing of the phantom divided barrier in a
Lorentz invariance violating dark energy model. Lorentz invariance
violation which is achieved by introducing a vector field in the
action, incorporates directly in the dynamics of the scalar field
and equation of state. This interesting feature allows us to study
phantom divided barrier crossing in the context of Lorentz
invariance violation. We show that for suitable choice of parameter
space, equation of state can cross phantom divided barrier just by
one scalar field and Lorentz violating vector field controls this
crossing.

\end{abstract}
\newpage

\section{Motivation}
Recently, Lorentz invariance violation (LIV) has been studied in the
context of scalar-vector-tensor theories[1]. It has been shown that
Lorentz violating vector fields affect the dynamics of the
inflationary models. One of the interesting feature of this scenario
is that the exact Lorentz violating inflationary solutions are
related to the absence of the inflaton potential. In this case, the
inflation is completely associated with the Lorentz violation and
depends on the value of the coupling parameter[2]. Standard
cosmology with a pressureless fluid as matter content of the
universe, predicts a universe either expanding forever or
re-collapsing eventually depending on the spatial geometry. Recent
evidences from supernova searches data [3,4], cosmic microwave
background (CMB) results [5, 6, 7] and also Wilkinson Microwave
Anisotropy Probe (WMAP) data [8, 9], indicate an accelerating phase
of cosmological expansion today and this feature shows that the
simple picture of universe consisting of pressureless fluid is not
enough; the universe may contain some sort of additional
negative-pressure dark energy. Analysis of the three year WMAP data
[10,11,12] shows that there is no indication for any significant
deviations from Gaussianity and adiabaticity of the CMB power
spectrum and therefore suggests that the universe is spatially flat
to within the limits of observational accuracy. Further, the
combined analysis of the three-year WMAP data with the supernova
Legacy survey (SNLS)[10], constrains the equation of state $w_{de}$,
corresponding to almost ${74\%}$ contribution of dark energy in the
currently accelerating universe, to be very close to that of the
cosmological constant value. Moreover, observations appear to favour
a dark energy equation of state, $w_{de}<-1$ [13]. Therefore a
viable cosmological model should admit a dynamical equation of state
that might have crossed the value $w_{de}= -1$, in the recent epoch
of cosmological evolution.  Various aspects of this crossing has
been studied extensively ( see for instance [14] and reference
therein). However, possible impact between LIV and phantom divided
barrier crossing has not been studied yet. Since there are some
traces of Lorentz invariance violation in high-energy regime
[15,16], it is interesting to study possible implication of this
symmetry breaking on the dynamics of equation of state and
especially crossing of the phantom divide barrier. The purpose of
this letter is to take a small step in this direction.
\section{A Lorentz Violating Cosmology}
In this section, following [1,2], we summarize the cosmological
dynamics of Lorentz invariance violating fields. Our goal is to find
a relation between Lorentz Invariance violation parameter and
dynamics of scalar field. This relation will affect the equation of
state of scalar field which is the central object of subsequent
sections.\\
We start with the following action for a typical
scalar-vector-tensor theory which admits Lorentz invariance
violation
\begin{eqnarray}
     S&=& S_g + S_u + S_{\phi} \ ,
     \end{eqnarray}
where the actions for the tensor field $S_g$, the vector field
$S_u$, and the scalar field $S_{\phi}$ are defined as follows
\begin{eqnarray}
        S_g &=& \int d^4 x \sqrt{-g}~ {1\over 16\pi G}R  \ ,
    \\
  S_u &=& \int d^4 x \sqrt{-g} \left[
  - \beta_1 \nabla^\mu u^\nu \nabla_\mu u_\nu
   -\beta_2 \nabla^\mu u^\nu \nabla_\nu u_\mu    -\beta_3 \left( \nabla_\mu u^\mu \right)^2 \right. \nonumber\\
 && \left.
  -\beta_4 u^\mu u^\nu \nabla_\mu u^\alpha \nabla_\nu u_\alpha
    + \lambda \left( u^\mu u_\mu +1 \right) \right]  \ ,
  \label{eq:act-VT} \\
  S_{\phi}  &=&  \int d^4 x \sqrt{-g}~ {\cal{L}}_{\phi}  \ .
   \end{eqnarray}
This action is allowed to contain any non-gravitational degrees of
freedom in the framework of Lorentz violating scalar-tensor-vector
theory of gravity. As usual, we assume $u^\mu u_\mu = -1$ and that
the expectation value of vector field $u^\mu$ is $<0| u^\mu u_\mu
|0> = -1$\,[17]. $\beta_i(\phi)$ ($i=1,2,3,4$) are arbitrary
parameters with dimension of mass squared and ${\cal{L}}_{\phi}$ is
the Lagrangian density for scalar field. Note that
$\sqrt{\beta_{i}}$ are mass scale of Lorentz symmetry
breakdown\,[1,17]. The detailed cosmological consequences of this
action are studied in Ref.[1].

Assuming a homogeneous and isotropic universe, we describe the
universe with the following metric
\begin{eqnarray}
ds^2 = - {\mathcal{N}}^2 (t) dt^2 + e^{2\alpha(t)} \delta_{ij} dx^i
dx^j \ ,
\end{eqnarray}
where ${\mathcal{N}}$ is a lapse function and the scale of the
universe is determined by $\alpha$\,[1,2]. By variation of the
action with respect to metric and choosing a suitable gauge, one
obtains the following field equations
\begin{eqnarray}
   R_{\mu\nu}-{1\over 2}g_{\mu\nu}R = 8\pi G T_{\mu\nu} \ ,
   \end{eqnarray}
where $T_{\mu\nu} =T_{\mu\nu}^{(u)} + T_{\mu\nu}^{(\phi)}$ is the
total energy-momentum tensor, $T_{\mu\nu}^{(u)}$ and
$T_{\mu\nu}^{(\phi)}$ are the energy-momentum tensors of vector and
scalar fields, respectively. The time and space components of the
total energy-momentum tensor are given by[2]
\begin{eqnarray}
     T^{0}_{0} = - \rho_u -\rho_{\phi} \ , \qquad    T^{i}_i =  p_u+ p_{\phi} \ ,
     \end{eqnarray}
where the energy density and pressure of the vector field are
calculated as follows
\begin{eqnarray}
    && \rho_u =  -3\beta H^2  \ ,
   \\
    &&p_u =  \left(3 + 2{H^{\prime}\over H} + 2{\beta^{\prime}\over \beta} \right)\beta H^2 \ ,
     \\
    && \beta \equiv \beta_1 +3 \beta_2 + \beta_3 \ ,
    \end{eqnarray}
a prime denotes the derivative of any quantity with respect to
$\alpha$ \, and \, $H\equiv d\alpha/dt=\dot{\alpha}$ is the Hubble
parameter. One can see that $\beta_4$ does not contribute to the
background dynamics[1,2]. The energy equations for the vector field
$u$ and scalar field, $\phi$ are as follows
\begin{eqnarray}
   {\rho}^{\prime}_u + 3({\rho}_u + p_u)=+3H^2 \beta^{\prime}  \ ,
   \end{eqnarray}
\begin{eqnarray}
    {\rho}^{\prime}_{\phi} + 3({\rho}_{\phi} + p_{\phi})=-3H^2 \beta^{\prime}  \
    ,
   \end{eqnarray}
respectively. So, the total energy equation in the presence of both
the vector and the scalar fields reads
\begin{equation}
   {\rho}^{\prime} + 3({\rho} + p)=0 \ , \quad (\rho = \rho_u + \rho_{\phi}) .
   \end{equation}
With these preliminaries, dynamics of the model is described by the
following Friedmann equations[1,2]
\begin{eqnarray}
     \left( 1 + \frac{1}{8\pi G \beta} \right) H^2={1\over 3\beta} \rho_{\phi}   \ ,
    \\
     \left( 1 + \frac{1}{8\pi G \beta} \right) \left( HH'+H^2\right)=-{1\over 6} \left( {\rho_{\phi}\over \beta} + {3p_{\phi}\over \beta} \right) - H^2 {\beta'\over \beta} \ .
    \end{eqnarray}
In the absence of vector field, that is, when  all \,$\beta_i =0$,
one recovers the standard equations of dynamics. For the scalar
sector of our model we assume the following Lagrangian
\begin{eqnarray}
{\cal{L}}_{\phi}= -{\eta\over 2}(\nabla \phi)^2 - V(\phi) \ ,
\end{eqnarray}
where $(\nabla
\phi)^2=g^{\mu\nu}\partial_{\mu}\phi\partial_{\nu}\phi$. Ordinary
scalar fields are correspond to $\eta = 1$ while $\eta = -1$
describes phantom fields. For the homogeneous scalar field, the
density $\rho_{\phi}$ and pressure $p_{\phi}$ are given as follows
\begin{eqnarray}
&&\rho_{\phi} = {\eta\over 2} H^2 \phi^{\prime 2} + V(\phi) \ ,
\\
&&p_{\phi}= {\eta\over 2} H^2 \phi^{\prime 2} - V(\phi) \ .
\end{eqnarray}
The corresponding equation of state parameter is
\begin{eqnarray}
\omega_{\phi}={p_{\phi}\over \rho_{\phi}} = - \frac{1- \eta H^2
\phi^{\prime 2}/2V}{1 + \eta H^2 \phi^{\prime 2}/2V} \ .
\end{eqnarray}
Now the Friedmann equation takes the following form \,[2]
\begin{eqnarray}
H^2  = \frac{1}{3\bar{\beta}} \left[
  \frac{\eta}{2} H^2 \phi^{\prime 2} + V(\phi) \right] ,
  \end{eqnarray}
where $\bar{\beta}=\beta+\frac{1}{8\pi G}$.  Using this equation we
can show that
\begin{eqnarray}
  \phi^{\prime} &=&-2\eta \bar{\beta}\left(\frac{H_{,\phi}}{H} +
\frac{\bar{\beta}_{,\phi}}{\bar{\beta}} \right) \ .
     \end{eqnarray}
Substituting this equation into the Friedmann equation, the
potential of the scalar field can be written as\,[2]
\begin{eqnarray}
    V = 3\bar{\beta} H^2 \left[ 1-{2\over 3}
    \eta\bar{\beta}\left({\bar{\beta}_{,\phi}\over \bar{\beta}}
    + {H_{,\phi}\over H}\right)^2 \right] \ .
\end{eqnarray}
Note that in the above equations the Hubble parameter $H$ has been
expressed as a function of $\phi$, $H=H(\phi(t))$. One can show that
the equation of state has the following form
\begin{eqnarray}
    \omega_\phi &=& -1 + {4\over 3}\eta\bar{\beta}\left(\frac{H_{,\phi}}{H} +
    \frac{\bar{\beta}_{,\phi}}{\bar{\beta}} \right)^2 \nonumber\\
    &=&-1 + {1\over 3}\eta \frac{\phi^{\prime 2}}{\bar{\beta}} \ .
    \end{eqnarray}
Equations (21) and (23) are essential in forthcoming discussions.
Note that violation of the Lorentz invariance which has been
introduced by existence of a vector field in the action, now has
incorporated in the dynamics of scalar field and equation of state
via existence of $\bar{\beta}$. This interesting feature allows us
to study phantom divided barrier crossing in the context of Lorentz
invariance violation. We need to solve these two equations, (21) and
(23),\,  to find dynamics of scalar field $\phi$ and the equation of
state $\omega_\phi$. This will be achieved only if the Hubble
parameter $H(\phi)$ and the vector field coupling,\,
${\bar{\beta}}(\phi)$ are known. In which follows, our strategy is
to choose some different cases of the Hubble parameter $H(\phi)$ and
the vector field coupling ${\bar{\beta}}(\phi)$ and then
investigating possible crossing of phantom divided barrier in this
context. We concentrate on suitable domains of parameter space which
admit such a crossing.

\section{LIV and Crossing of the Phantom Divided Barrier}
To investigate phantom divided barrier crossing in this context, we
have to solve equations (21) and (23) for four unknowns: \,$H$,
$\omega_\phi$, $\bar{\beta}$, and $V$. This is impossible unless two
of these unknowns be specified a priori. A large class of equations
of state for scalar field has been studied in [18] and [19] and some
classes of potentials allowing for the scalar field equation of
state were described. Also some authors have used vector field
models for coincidence of dark energy[20,21,22,23]. The main point
of these studies which is crucial for our subsequent discussions is
the fact that to have a successful model with phantom divided
barrier crossing within a minimally coupled scalar field scenario,
one should consider both quintessence and phantom fields. However,
with non-minimally coupled scalar fields, one can achieve phantom
divided barrier crossing just with one of these fields[14]. As we
will show, in the presence of Lorentz invariance violating fields,
it is possible to cross phantom divided barrier just by considering
one field, i. e., quintessence field. This may reflect the fact that
LIV has something to do with non-minimal coupling of scalar field
and gravity. In other words, our basic action defined in equations
(1) and (2) are actually minimally coupled, but our results for
crossing of phantom divide barrier are very similar to non-minimal
model results presented in [14]. So, it seems that there is a close
relation between LIV and non-minimal scenario in the present
context. In which follows, based on above argument, we consider just
a quintessence scalar field with $\eta=1$. We study possible
crossing of the phantom divided barrier in some model universes with
concentration on suitable range of the parameter space.\\

\subsection {Case 1}
In the first step we consider a model universe with the following
simple choices
\begin{equation}
    H=H_0 \ , \qquad  \bar{\beta}(\phi) = m\phi^2,
    \end{equation}
where $H_0$ is a positive constant. With these choices, equations
(21) and (23) lead us to the following relations
\begin{equation}
   \phi(t)=\phi_0 \exp \left[-4\eta mH_0(t-t_0) \right] \ ,
   \end{equation}
\begin{eqnarray}
   \omega_\phi &=& -1 + {16\over 3} m,
       \end{eqnarray}

where $\phi(t=t_0)\equiv \phi_0$ is a constant. This model shows
that cosmic evolution starts from a constant value of the scale
factor and then grows exponentially, $a(t)=a_0 e^{H_0(t-t_0)}$.
Equation (26) shows that the equation of state $\omega_\phi$ has no
dynamics since it only depends on the value of the vector field
coupling parameter $m$. Since an accelerated expansion occurs for
$\omega_\phi < -1/3$ [2], then we should have $m<1/8$ for a typical
quintessence field. However, the present data of the universe seems
to tell us that $\omega_\phi$ might be less than $-1$. Thus, the
value of $m$ may be chosen to fit with the present observable
constraint on the equation of state parameter. If we use the values
of $m$ obtained in [2], we can calculate $\omega_\phi$ versus $m$.
This is shown in figure 1.\
\begin{figure}[htp]
\begin{center}\includegraphics{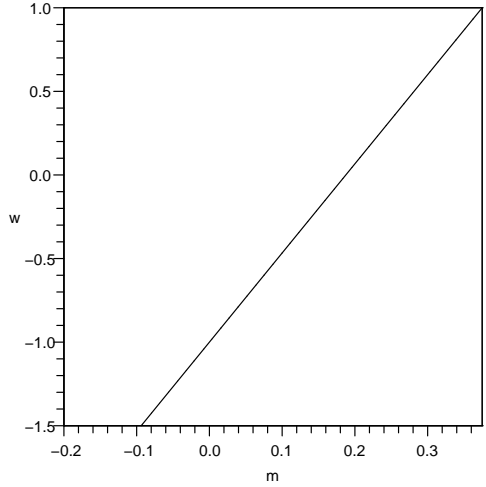} \vspace{6cm}
\end{center}
 \caption{\small {Variation of $\omega_\phi$
for different values of the coupling vector parameter $m$. }}
\end{figure}
This figure shows that for crossing of phantom divided line,
$\omega_\phi=-1$, the lower value of $m$ should be negative. However
this is impossible since $m$ is related to $\sqrt{\beta_{i}}$ as
mass scales of Lorentz symmetry breakdown. So, in this model
universe, crossing of phantom divided barrier with one scalar field
is impossible. This is not surprising since equation of state in
this case has no dynamics and there is no reasonable fine tuning.
However it is simple to show that existence of two scalar fields,
one quintessence and the other phantom field, as usual can lead to
possibility of cosmological constant equation of state crossing.
Note that in Ref.[2], the authors have found that $m$ should be
restricted to interval $1/6<m<3/8$ to provide a suitable inflation
model. In summary, this model universe provides no consistent
scenario for a dynamical equation of state containing phantom
divided barrier crossing.

\subsection{Case\,2}
As a second case, we suppose \, $H(\phi)=H_0\phi^\xi$\, and \,
$\bar{\beta}(\phi)=m\phi^2$. With these choices, we find the
following equation of state
\begin{eqnarray}
   \omega_\phi = -1 + {4\over 3} m (\xi + 2)^2 \ .
   \end{eqnarray}
which evidently has no dynamics. The condition for acceleration of
the universe, $H'/H > -1$,  yields
\begin{eqnarray}
   m < {1\over 2\xi (\xi + 2)} \ .
   \end{eqnarray}
\begin{figure}[htp]
\begin{center}\includegraphics{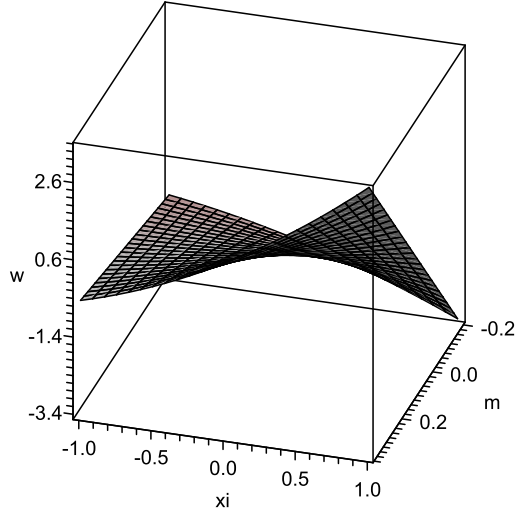} \vspace{6cm}
\end{center}
 \caption{\small {Variation of $\omega_\phi$
for different values of the vector field coupling parameter $m$ and
$\xi$. Note that based on relation (28), if we choose $-1<\xi<1$
then $m$ should be restricted to the interval $-0.5<m<0.166$.
However $m$ can not attain negative values and therefore we should
restrict this interval to $0<m<0.166$. }}
\end{figure}
The scalar field evolve as follows
\begin{equation}
   \phi(t)=\phi_0\left(1+{H_0\phi_0^\xi\over p}(t-t_0)\right)^{-1/\xi} \ .
   \end{equation}
The variation of $\omega_\phi$ versus $m$ and $\xi$ is shown in
figure $2$. This figure shows that only for positive values of $m$
and all possible values of $\xi$,\, the equation of state satisfies
the condition $\omega_{\phi} >-1$. For other positive values of
$\xi$ and negative values of $m$, equation of state has a phantom
divide line since $\omega_{\phi} < -1$. But we should stress that
since $m$ cannot attain negative values, we must restrict $m$ to the
interval $0<m<0.166$. As figure shows, with this restriction it is
impossible to cross phantom divided line with one scalar field.\\

In these two examples, the equation of state has no dynamics. Only
with variation of parameters $m$ and $\xi$ one can formally attain a
phantom divided barrier crossing. The case with a dynamical equation
of state is more reasonable. Now we turn to the situation where the
equation of state is dynamical. For this purpose, we generalize the
vector field coupling to achieve the value of
$\bar{\beta}(\phi)=m\phi^n$,\, for\, $n>2$.

\subsection{Case\,3 }
To have a dynamical equation of state, we consider a model universe
where the vector field coupling parameter is written as a power of
the scalar field
\begin{equation}
    H=H_0 \ , \quad  \bar{\beta}(\phi) = m\phi^n \ ,\quad n >2 \ ,
    \end{equation}
where $H_0$ and $n$ are constant positive parameters. Following the
above procedure, the scalar field $\phi$, the vector field coupling
$\bar{\beta}$ and equation of state $\omega_{\phi}$ have the
following dynamics[2]
\begin{equation}
   \phi(t)=\frac{\phi_0}{ \left[1 + 2mnH_0(n-2)\phi_0^{n-2}(t-t_0) \right]^{1\over{n-2}}},
\end{equation}
\begin{equation}
   \bar{\beta}(t)=\frac{m\phi_0^n}{ \left[1 + 2mnH_0(n-2)\phi_0^{n-2}(t-t_0) \right]^{n\over{n-2}}},
\end{equation}
\begin{eqnarray}
  \omega_{\phi}(t) = -1 + \frac{4mn^2\phi_0^{n-2}/3}{1+2 mnH_0(n-2)\phi_0^{n-2}(t-t_0)} \ .
  \end{eqnarray}
Remember that $\bar{\beta}(t)$ plays the role of Lorentz invariance
violation in this setup. Equation of dynamics for $\bar{\beta}(t)$
implicitly has an important meaning: by a suitable fine tuning one
can construct a Lorentz violating cosmology consistent with
observational data. In another words, this setup provides an
important basis for testing LIV in cosmological context.\\
Although many different models can also lead to phantom divided
barrier crossing, our model is special in this respect since it
contains only one scalar field and the presence of Lorentz violating
vector field controls the crossing. In this sense, fine tuning of
parameters space based on observational data restricts the value
that $\bar{\beta}(t)$ can attain. Any non-vanishing value of
$\bar{\beta}$ in our model shows violation of Lorentz symmetry in
this cosmological setup. If dynamics of $\bar{\beta}(t)$ which is
described by equation (32) can be detected and constraint by
observational data, this will be a manifestation of LIV in
cosmological context. Lorentz invariance violating inflation models
constraint by WMAP and other observational data my provide other
test of LIV in cosmological setup. To see possible detection of
Lorentz-violating field in cosmology see [25,28,29].

\begin{figure}[htp]
\begin{center}\includegraphics{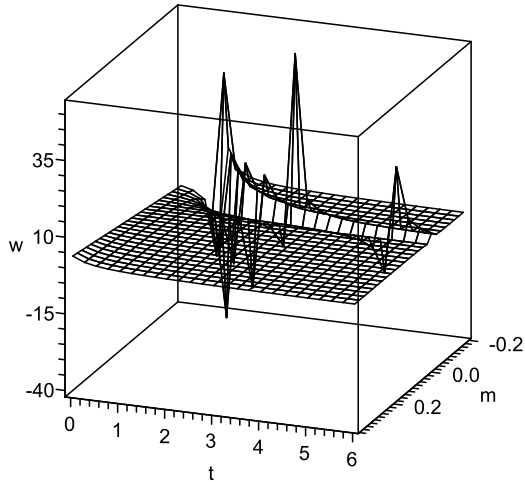} \vspace{6cm}
\end{center}
 \caption{\small {Variation of $\omega_\phi$
for different values of the vector field coupling $m$ and $t$ for
$n=3$. }} \end{figure}

In this model, the cosmic evolution starts from a constant value of
the scale factor and then grows exponentially, $a(t)=a_0
e^{H_0(t-t_0)}$, while the vector field coupling, $\bar{\beta}$,
starts from a constant value of the scalar field, $m\phi_0^n$. The
equation of state $\omega_\phi$ is dynamical. Figure $3$ shows the
variation of the equation of state versus $m$ and $t$. This figure
clearly demonstrates a crossing of the phantom divided line. We
stress that such a results are essentially model dependent. In
principle, these results should be translated to red-shift language
which provides a better framework to compare with observational
data. While this is essentially important, the purpose of this
letter is to show that LIV may help us to achieve phantom divided
line crossing with only one scalar field. In other words in the
presence of LIV, just one scalar field is enough to achieve phantom
divided barrier crossing and existence of vector field controls the
situation.
\subsection{A more General case }
In this section we consider a more general case where both the
vector field coupling and the Hubble parameter are functions of
$\phi$ defined as follows
\begin{equation}
    H=H_0\phi^{\xi} \ , \quad  \bar{\beta}(\phi) = m\phi^n \ ,\quad n >2 \
        \end{equation}
Using equation (21), for this case we obtain
\begin{equation}
\phi \left( t \right) = \frac{1}{\Big[H_0(t-t_0)(-4\,\xi m+4\,\xi
mn+2\,{\xi}^{2}m-4 \,mn+2\,m{n}^{2})+\phi_0 \Big ] ^{
\left(\frac{1}{ n+\xi-2} \right) }}
\end{equation}
and using equation (23) we find
\begin{equation}
\omega_\phi(t)=-1+\frac{4}{3}m\phi^{n-2}(t)(\xi+n)^2
\end{equation}
The condition for acceleration of the universe, that is, $H'/H >
-1$, yields
\begin{equation}
m^2<\frac{1}{4(-1)^n\phi^{n-2}(t)(\xi+n)^2},~~~~~~ n>2 ,
\end{equation}
With $\phi$ defined as (35), the equation of state takes the
following form
\begin{equation}
\omega_\phi(t)=-1+\frac{4}{3}m\frac{(\xi+n)^2}{\Bigg[H_0(t-t_0)(-4\,\xi
m+4\,\xi mn+2\,{\xi}^{2}m-4 \,mn+2\,m{n}^{2})+\phi_0 \Bigg] ^{
\left(\frac{n-2}{ n+\xi-2} \right) }},
\end{equation}
which explicitly has a dynamical behavior. This model allows us to
choose a suitable parameter space to cross phantom divided barrier.
This parameter space should be checked by observational data to have
a reasonable model. The most important aspect of the present model
is the fact that in principle, LIV fields provide situation that one
scalar field and another vector field together can lead us to
describe phantom divided barrier crossing.
\begin{figure}[htp]
\begin{center}\includegraphics{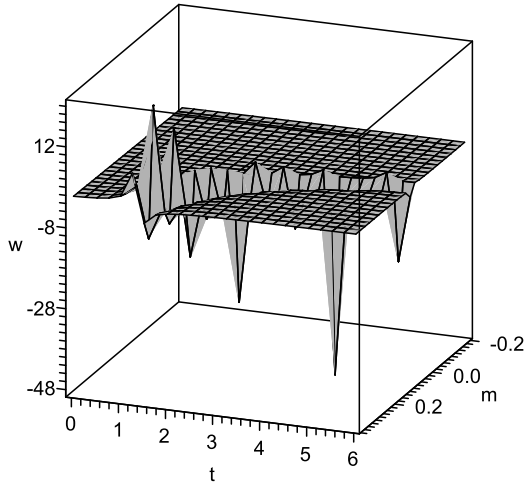} \vspace{6cm}
\end{center}
 \caption{\small {Variation of $\omega_\phi$
for different values of the vector field coupling $m$ and $t$ for
$n=3$ and $\xi=-2$. Positive values of $\xi$ show no phantom divided
barrier crossing. The values of $\xi$  are determined by relation
(37). }}
\end{figure}

Figure $4$ shows the crossing of phantom divided barrier for a
dynamical equation of state. Figure $4$ may be used to explain why
we are living in an epoch  of \, $\omega< -1$ since in late time we
see that $\omega< -1$. This is the second cosmological coincidence
problem. Two point should be stressed here: firstly, as figures $3$
and $4$ show, there are some sudden jumps of the equation of state.
In many existing models whose equation of state can cross the
phantom divided barrier, $\omega$ undulates around $-1$ randomly
([24] and references therein). These jumps are actually a
manifestation of this undulation which may be a signature of chaotic
behavior of equation of state. Secondly, as these figures show,
crossing of the phantom divided barrier can occur at late-time. This
fact, as second cosmological coincidence problem, needs additional
fine-tuning in model parameters and trigger mechanism, for instance,
can be used to alleviate this coincidence.

\section{Conclusions}
As a new mechanism for crossing of phantom divided barrier by
equation of state, in this paper we have incorporated a possible
violation of Lorentz invariance in a cosmological setup. We have
shown that by a suitable choice of parameter space, it is possible
to have phantom divided barrier crossing in a Lorentz invariance
violating context just by a single scalar field. This is an
important result since in the absence of LIV, as previous studies
have shown, it is impossible to cross phantom divided barrier by
just one scalar field minimally coupled to gravity[14,27]. In this
regard, existence of a Lorentz invariance violating vector field
provides a framework for crossing phantom divided barrier with one
scalar field. On the other hand, this model provides a possible
framework for testing Lorentz invariance violation in a cosmological
context. Using observational data and by a suitable fine tuning it
is possible to construct a reliable Lorentz violating cosmological
model. A similar strategy has been applied for inflation in
reference[2]. Another aspect of our model is the fact that it
contains several crossing of phantom divided barrier, a phenomena
which has been seen in other scenarios[30,31,32]. As we have shown,
the equation of state takes different form in different choices of
parameter space.\\
In this framework it is possible to use the "trigger mechanism" to
explain dynamical equation of state. This means that we assume
scalar- vector-tensor theory containing Lorentz invariance violation
which acts like the hybrid inflation models. In this situation,
vector and scaler field play the roles of inflaton and the
"waterfall" field respectively. This conjecture is under
investigation[26].\\

\end{document}